# Windowed Catodyne Measurements: Continuum Approximation versus Exact Discrete Photon-Number Treatment

Elahe Nahvifard[1], Mohammad Reza Bazrafkan[1]

[1] *Physics Department, Faculty of Science, Imam Khomeini International University, Qazvin 34149-16818, I. R. Iran*

**Abstract**

**We analyze a family of measurements obtained by retaining only a finite window of the photon-sum variable in catodyne detection, where a Schrödinger cat state serves as the local oscillator. In the continuum approximation, the window half-width $\delta$ appears to act as a continuous knob for coherence. However, a treatment on the discrete photon-number lattice shows that, in the strong-LO regime, the coherence conditioned on the difference outcome $x$ is not suppressed by the window. The exact parity-sensitive success probability is a staircase function of $\delta$ and $|\beta|$, whose plateau structure and magnitude the continuum formulas fail to capture, especially for small $|\beta|$ and narrow windows. For remote preparation of cat states, an exact calculation with a finitely squeezed resource shows that the remote state keeps a purity close to one when x is resolved, and loses purity mainly when x is not recorded. Our results clarify where the continuum approximation of catodyne detection can be trusted and where the discrete photon-number structure must be used.**

*nahvifard@sci.ikiu.ac.ir*

## 1 Introduction

Quantum-optical measurements are the backbone of modern quantum technologies, from quantum state tomography to continuous-variable quantum computing. Among these, balanced homodyne detection is the most widely used technique for measuring field quadratures [1, 2]. In the standard configuration, a signal mode is mixed with a strong coherent-state local oscillator (LO) at a 50:50 beam splitter, and the photocurrent difference yields a quadrature observable in the limit of large LO amplitude [1, 3]. The corresponding positive operator-valued measure (POVM) is a rank-1 projector onto a quadrature eigenstate $|x_\theta\rangle$, where $\theta$ is the LO phase [4].

The LO, however, need not be a coherent state. Nonclassical LO states can give rise to nonclassical measurements with properties beyond standard homodyne detection [5, 6, 7]. A particularly interesting case is the use of a Schrödinger-cat state $|\beta\rangle \pm |-\beta\rangle$ as the LO, a scheme referred to as "catodyne" [8]. Combes and Lund derived the exact Kraus operators and POVM elements for catodyne and showed that, in the strong-LO limit, the POVM elements contain both diagonal terms and coherence terms of the form $|x_\theta\rangle\langle -x_\theta|$ [8]. They identified two limiting cases. If both the difference and the sum of the photon counts are retained, the POVM has coherence between $|x_\theta\rangle$ and $\langle -x_\theta|$ (rank 1). If the sum is integrated out in the continuum approximation, this coherence vanishes and the measurement becomes reflection-symmetric (rank 2). They also noted that the ability to use or ignore the information in the sum variable may provide a means to engineer a measurement [8].

This observation raises a natural question that, to our knowledge, has not been studied quantitatively: what happens between the two limits, when only a finite window of the sum variable is retained? In the continuum approximation, where the sum variable $w$ is treated as a continuous Gaussian-distributed quantity, the window half-width $\delta$ appears to interpolate smoothly between the two limits and to act as a continuous tuning knob for coherence. A related line of work has shown how to tune between photon-number and quadrature measurements by varying the amplitude of a coherent-state LO [9]. Here the LO amplitude is fixed, and the parameter under study is a property of how the photon-counting record is processed.

However, the continuum picture is only an approximation. In an actual experiment the photon counts $n$ and $m$ are integers, so that and $\sqrt{2}|\beta|x = n-m$ are integers of equal parity. In this work we a $\sqrt{2}|\beta|w = n+m$ analyze the windowed catodyne measurement both in the continuum approximation and on the exact discrete photon-number lattice, and we find that the two descriptions differ qualitatively.

The main contributions of this work are threefold:

**1-Coherence conditioned on x is preserved in the strong-LO regime.** We show that on the discrete photon-number lattice, windowing the sum variable $w$ does not reduce the coherence conditioned on the difference outcome$x$. The apparent suppression of coherence in the continuum model is an artifact of the continuum approximation. Coherence is lost only when $x$ is also integrated out or ignored.

**2-Parity-sensitive acceptance.** We derive exact windowed success probabilities for Fock-state inputs and define a parity-discrimination signal D. We show that D is a staircase function of δ and |β|, with magnitudes and

plateaus that differ substantially from the continuum prediction.

**3-Limits of the continuum approximation.** By comparing exact discrete results with the continuum formulas, we quantify where the continuum approximation breaks down. The breakdown is strongest for small $|\beta|$ and narrow windows, where the window contains only a few photon-number values, as is the case for realistic photon-number-resolving detectors (see Sec. 6.4).

The structure of this paper is as follows. In Sec. 2 we introduce the setup and notation. In Sec. 3 we define the windowed POVM, derive its continuum-approximation expressions, and then present the exact discrete treatment. In Sec. 4 we present numerical results comparing the two. In Sec. 5 we discuss the discrete parity structure. In Sec. 6 we discuss implications for measurement engineering and remote state preparation. We conclude in Sec. 7.

## 2 Setup and Notation

We consider a balanced homodyne setup in which a signal mode $a$ and a LO mode $b$ are mixed at a 50:50 beam splitter. The output modes are

$$c = \frac{a+b}{\sqrt{2}}, \quad d = \frac{a-b}{\sqrt{2}}, \tag{1}$$

and the two output ports are measured by ideal photon-number-resolving detectors [8]. The joint input state is $\rho_{in} = \rho \otimes |\text{LO}\rangle\langle\text{LO}|$, where the signal state is. For catodyne, the LO is prepared in a cat state

$$\begin{aligned} |\text{cat}_\pm(\beta)\rangle &= \frac{1}{\mathcal{N}_\pm}\left(|\beta\rangle \pm |-\beta\rangle\right), \\ \mathcal{N}_\pm &= \sqrt{2\left(1 \pm e^{-2|\beta|^2}\right)}. \end{aligned} \tag{2}$$

where $\beta = |\beta| e^{i\theta}$ is the complex amplitude of the coherent state, $|\beta|$ is the LO amplitude, $\theta$ is the LO phase, and $\mathcal{N}_\pm$ is the normalization constant [8]. The upper sign corresponds to the even cat and the lower sign to the odd cat.

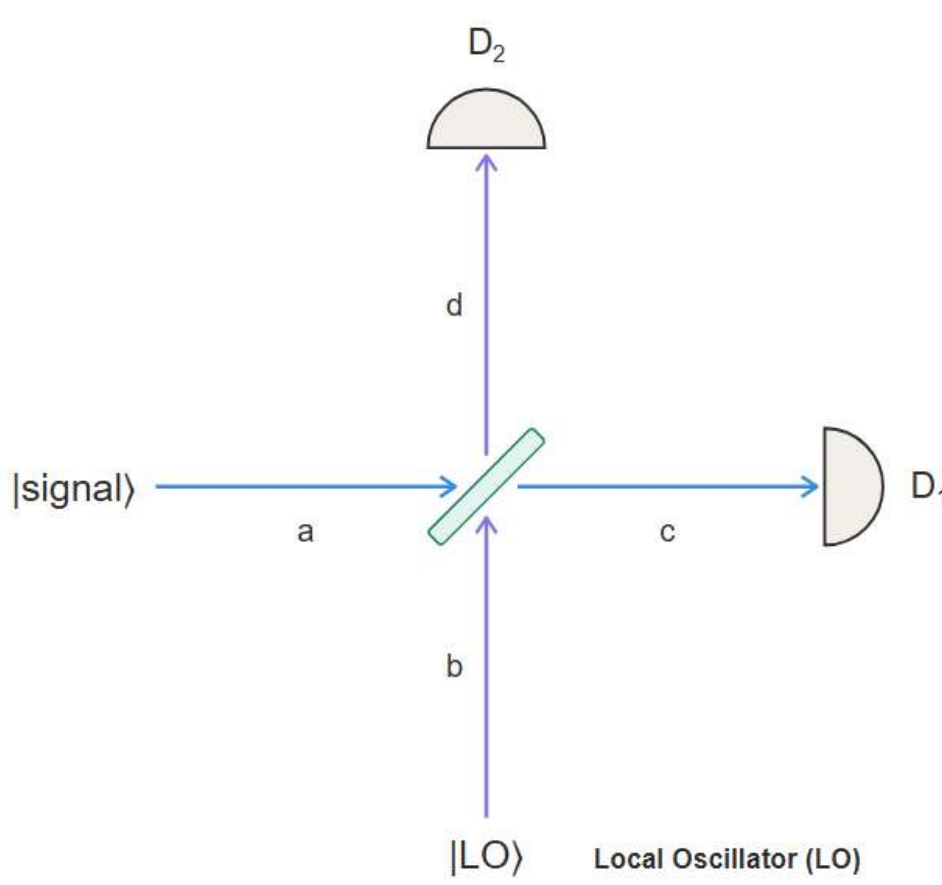

**Fig. 1**. Balanced homodyne setup. A signal state $|\text{signal}\rangle$ is mixed with a local oscillator $|\text{LO}\rangle$ at a 50:50 beam splitter. The two output ports $\text{c}$ and $\text{d}$ are measured by ideal photon-number-resolving detectors $\text{D}_1$ and $\text{D}_2$. Detector 1 registers $n$ photons and detector 2 registers $m$ photons.

The Kraus operator for detecting $n$ photons in mode $c$ and $m$ photons in mode $d$ is obtained by projecting the output state onto the corresponding Fock states [8]

$$\begin{aligned} M^{[\beta]}_{n,m} &= \langle 0| \frac{e^{-|\beta|^2/2}}{2^{(n+m)/2}\sqrt{n!m!}\,\mathcal{N}_\pm(\beta)} \times \\ &\left[(a+\beta)^n (a-\beta)^m \pm (a-\beta)^n (a+\beta)^m\right]. \end{aligned} \tag{3}$$

This expression is valid for all $\beta$ and acts only on the signal mode [8]. The two terms correspond to the two coherent components of the cat state and are added at the amplitude level, which is the origin of the quantum coherence in catodyne.

We introduce the scaled variables

$$x = \frac{n-m}{\sqrt{2}|\beta|}, \qquad w = \frac{n+m}{\sqrt{2}|\beta|}. \tag{4}$$

Because $n$ and $m$ are integers, $\sqrt{2}|\beta|x = n-m$ and $\sqrt{2}|\beta|w = n+m$ are integers of equal parity, so $x$ and $w$ take values on a discrete lattice of spacing $1/\left(\sqrt{2}|\beta|\right)$. The strong-LO expressions below hold when $|\beta|^2$ is much larger than $\left|\langle a^4\rangle\right|$ [1, 8]; for small $|\beta|$ they are only approximate, and we compare them with the exact Kraus operator of Eq. (3) in Secs. 3.4, 5 and 6.3.2.

In the strong-LO limit, the Kraus operator factorizes as [8]

$$\begin{aligned} &M^{[\beta]\pm}_{x,w}\sqrt{dwdx} \approx \frac{1}{\sqrt{2}} e^{i\sqrt{2}|\beta|w\theta} \times \\ &\left[\langle x_\theta| e^{i\pi\sqrt{2}|\beta|(x-w)/2} \pm \langle -x_\theta| e^{i\pi\sqrt{2}|\beta|(x+w)/2}\right] \times \\ &\frac{e^{-\left(w-|\beta|/\sqrt{2}\right)^2/2}}{\pi^{1/4}}\sqrt{dwdx}. \end{aligned} \tag{5}$$

This factorization shows that the Kraus operator decomposes into a combination of quadrature eigenstates and a Gaussian weight in $w$. The corresponding POVM element, obtained by squaring this expression, is

$$\begin{aligned} &dwdxE^{\pm[\beta]}_{x,w} \approx dwdx \frac{e^{-\left(w-|\beta|/\sqrt{2}\right)^2/2}}{\sqrt{\pi}} \frac{1}{2} \times \\ &\Big[|x_\theta\rangle\langle x_\theta| + |-x_\theta\rangle\langle -x_\theta| \pm \\ &|x_\theta\rangle\langle -x_\theta| e^{\sqrt{2}i\pi|\beta|w} \pm |-x_\theta\rangle\langle x_\theta| e^{-\sqrt{2}i\pi|\beta|w}\Big]. \end{aligned} \tag{6}$$

The first two terms are diagonal and describe the classical part of the measurement. The last two terms are coherence terms that couple the quadrature eigenstates and are responsible for the nonclassical character of the measurement [8].

## 3 Windowed Sum Variable: Continuum Approximation and Exact Treatment

### 3.1 Definition

We define the partial-resolution POVM by integrating $E_{x,w}^{|\beta|\pm}$ over a finite window of $w$:

$$E_{x,w}^{(\delta,\beta)\pm} \approx \int_{w_0-\delta}^{w_0+\delta} dw E_{x,w}^{\pm[\beta]}, \qquad w_0 = \frac{|\beta|}{\sqrt{2}}. \tag{7}$$

The window half-width $\delta$ is measured in units of $w$; in photon-number units it corresponds to $\sqrt{2}\,|\beta|\,\delta$.

This definition is the central concept of our analysis. In contrast to previous measurements, which either fully discard $w$ (no coherence) or fully retain it (full coherence), this definition generates a family of measurements parametrized by $\delta$.

### 3.2 Analytical Expressions

The following expressions treat $w$ as continuous. As shown in Sec. 3.5, this approximation misrepresents the coherence on the true lattice and is retained only as a large-$|\beta|$ reference.

Performing the integration, we obtain

$$E_x^{(\delta,\beta)} = \frac{G_\delta}{2}\left(|x_\theta\rangle\langle x_\theta| + |-x_\theta\rangle\langle -x_\theta|\right) \pm\frac{1}{2}\left[I_\delta(\beta)|x_\theta\rangle\langle -x_\theta| e \pm I_\delta^*(\beta)|-x_\theta\rangle\langle x_\theta|\right]. \tag{8}$$

where $G_\delta$ is the"success probability" (the fraction of the $w$ distribution that falls inside the window),

$$G_\delta = \int_{w_0-\delta}^{w_0+\delta} \frac{e^{-(w-w_\circ)^2}}{\sqrt{\pi}} dw = erf(\delta), \tag{9}$$

and

$$I_\delta = \int_{w_0-\delta}^{w_0+\delta} \frac{e^{-(w-w_\circ)^2}}{\sqrt{\pi}} e^{i\sqrt{2\pi}|\beta|w} dw, \tag{10}$$

is the "coherence weight" (how much coherence remains in the window). For the symmetric window, $I_\delta$ can be simplified by introducing the variable $w = w_\circ - y$.

$$I_\delta(\beta) = \frac{2e^{i\pi|\beta|^2}}{\sqrt{\pi}} \int_0^\delta e^{-y^2} \cos\left(\sqrt{2\pi}|\beta|y\right) dy, \tag{11}$$

The integral is real; the phase $I_\delta$ comes entirely from the global factor $e^{i\pi|\beta|^2}$, up to an additional phase of $\pi$ whenever the integral changes sign.

### 3.3 Normalized coherence factor

To quantify the degree of coherence retained, we define the "normalized coherence factor":

$$C_\delta(\beta) = \frac{|I_\delta(\beta)|}{G_\delta(\beta)} \quad . \tag{12}$$

This quantity is the ratio of the coherence weight to the success probability. To see why $0 \le G_\delta(\beta) \le 1$, we examine the two weights: $G_\delta$ is a real, positive integral of a Gaussian function, while $I_\delta$ is a complex integral of the same Gaussian with a phase factor. Since the modulus of an integral of a complex function cannot exceed the integral of its modulus, we have

$$|I_\delta| = \left|\int_{w_0-\delta}^{w_0+\delta} \frac{e^{-(w-w_\circ)^2}}{\sqrt{\pi}} e^{i\sqrt{2\pi}|\beta|w} dw\right| \le \int_{w_0-\delta}^{w_0+\delta} \frac{e^{-(w-w_\circ)^2}}{\sqrt{\pi}} \left|e^{i\sqrt{2\pi}|\beta|w}\right| dw = \int_{w_0-\delta}^{w_0+\delta} \frac{e^{-(w-w_\circ)^2}}{\sqrt{\pi}} dw = G_\delta. \tag{13}$$

Since $\left|e^{i\varphi}\right| = 1$ for any real phase $\varphi$, it follows that $|I_\delta| \le G_\delta$, and hence $0 \le C_\delta \le 1$. (The bound $0 \le G_\delta \le 1$ follows directly from $G_\delta = erf(\delta)$). Physical interpretation:" $C_\delta$ is the ratio of "retained coherence" to "total success probability." If $C_\delta = 1$, all accepted events have full coherence (a pure state). If $C_\delta = 0$, no coherence remains (a classical mixture).

In the limit of a narrow window $(\delta \to 0)$, we find $|I_\delta| \approx \frac{2\delta}{\sqrt{\pi}}$ and $G_\delta \approx \frac{2\delta}{\sqrt{\pi}}$, so that $C_\delta \to 1$. In the limit of a wide window $(\delta \to \infty)$, the oscillatory factor $\cos\left(\sqrt{2\pi}|\beta|y\right)$ causes the integral to cancel, so that $|I_\delta| \to 0$ and $C_\delta \to 0$ (up to a residual of order $e^{-\frac{\pi^2|\beta|^2}{2}}$).

Within the continuum approximation, $C_\delta$ therefore interpolates between 1 and 0. Sec. 3.5 shows that this behavior does not survive on the discrete photon-number lattice.

### 3.4 Parity-Discrimination Signal

The coherence weight $I_\delta$ also controls the "parity sensitivity" of the measurement. For a Fock state $|N\rangle$, the parity operator $P = (-1)^{a^\dagger a}$ acts as $P|N\rangle = (-1)^N |N\rangle$

In the continuum approximation, the success probability for detecting $|N\rangle$ in the window is

$$p_{\text{succ}}(N,\pm) = G_\delta \pm (-1)^N \operatorname{Re}(I_\delta) \tag{14}$$

because $\int dx\,|x\rangle\langle -x| = P$ The success probability thus depends on the parity of the input state, a parity dependence that is absent in standard homodyne detection. For the even cat, the continuum parity signal is the difference between even and odd inputs:

$$D_{\rm cont} = 2\,{\rm Re}\, I_\delta. \tag{15}$$

These expressions treat $w$ as continuous. On the true photon-number lattice, $n$ and $m$ are integers, so the window contains only the integer values $s = n + m$ satisfying

$$|\beta|^2 - \sqrt{2}\,|\beta|\,\delta \le s \le |\beta|^2 + \sqrt{2}\,|\beta|\,\delta. \tag{16}$$

These integers are the points of the window $[w_\circ - \delta, w_\circ + \delta]$ of Eq. (7), where $w = s/(\sqrt{2}\,|\beta|)$ and $w_\circ = |\beta|/\sqrt{2}$. We write $W$ for this window, so that $s \in W$ means that $w = s/(\sqrt{2}\,|\beta|)$ lies inside it. For a Fock input $|N\rangle$, the exact windowed success probability is

$$P(N,\pm) = \sum_{s\in w, n+m=s} e^{-|\beta|^2} \times \frac{\left|\langle 0|(a+\beta)^n (a-\beta)^m \pm (a-\beta)^n (a+\beta)^m\right|^2}{2^{(n+m)} n!\, m!\, \mathcal{N}_\pm^2(\beta)}. \tag{17}$$

where $\mathcal{N}_\pm^2 = 2\left(1 \pm e^{-2|\beta|^2}\right)$. and the matrix element is evaluated with Eq. (G4) of Combes and Lund [8]. We define the exact parity signal as

$$D_{\rm disc} = P(0,+) - P(1,+). \tag{18}$$

Both signals can take either sign, but the exact signal is piecewise constant and can be larger in magnitude. For example, for $|\beta| = 2$ and $\delta = 0.05$ we find $D_{\rm disc} = 0.39$, whereas the continuum formula gives $0.11$. For $|\beta| = 2.3$ and $\delta = 0.1$ we find $D_{\rm disc} = -0.33$.

For these two inputs the sums can be done in closed form. For the vacuum and an even cat, only even $s$ contribute, with $P(0,+) = \sum_{s\,{\rm even}} 4e^{-|\beta|^2} |\beta|^{2s}/(s!\,\mathcal{N}_+)$. For a one-photon input, only odd $s$ contribute, with $P(1,+) = \sum_{s\,{\rm odd}} 4e^{-|\beta|^2} |\beta|^{2(s-1)}/((s-1)!\,\mathcal{N}_+)$. Both sums run over $s \in w$.

### 3.5 Discrete-lattice treatment of the coherence (strong-LO)

We work in the strong-LO regime of Combes and Lund [8], where the Kraus operator for the outcome $(x, w)$ is, up to a global phase $e^{i\varphi(x,w)}$ and the Gaussian amplitude $\sqrt{g(w)}$,

$$M_{x,w} \propto \langle x| \pm e^{i\pi\sqrt{2}|\beta|w} \langle -x|. \tag{19}$$

Because $n$ and $m$ are integers, $\sqrt{2}\,|\beta|\,w = n + m$ and $\sqrt{2}\,|\beta|\,x = n - m$ are integers of equal parity. Hence

$$e^{i\pi\sqrt{2}|\beta|w} = (-1)^{n+m} = (-1)^{n-m} = (-1)^{\sqrt{2}|\beta|x}. \tag{20}$$

The relative sign between the two quadrature eigenstates therefore depends on $x$ only. It is identical for every value of $w$ that is compatible with a given $x$.

**Proposition.** On the discrete lattice, all outcomes $(x, w)$ with the same $x$ have Kraus operators that are proportional to each other. For a window $W = [w_\circ - \delta, w_\circ + \delta]$ of the sum variable, the windowed POVM element is

$$\begin{aligned} E_x^w &= \frac{G_x^w}{2}\left(|x\rangle + \sigma_x |-x\rangle\right)\left(\langle x| + \sigma_x \langle -x|\right), \\ \sigma_x &= \pm(-1)^{\sqrt{2}|\beta|x}, \end{aligned} \tag{21}$$

where $G_x^w$ is the sum of the Gaussian weights $g(w)$ over the lattice points of $w$ compatible with $x$. This is a rank-1 projector onto a cat state of definite parity. Consequently the normalized coherence factor is

$$C_\delta^{\rm disc}(x) = \frac{|I_x^w|}{G_x^w} = 1. \tag{22}$$

This holds for every window and every $\delta$.

**Proof.** The Kraus operators for fixed $x$ differ only by a scalar factor $\sqrt{g(w)}e^{i\varphi(x,w)}$, since the relative sign $\sigma_x$ is the same for all compatible $w$. Summing $E_{x,w} = M_{x,w}^\dagger M_{x,w}$ over $w \in W$ therefore multiplies the rank-1 operator $\left(|x\rangle + \sigma_x |-x\rangle\right)\left(\langle x| + \sigma_x \langle -x|\right)/2$, weight $G_x^w$. The coherence term carries the same weight as the diagonal terms, so $|I_x^w| = G_x^w$.

**Consequences.**

1-Windowing $w$ is a coarse-graining of outcomes whose Kraus operators are proportional. Such a coarse-graining cannot reduce coherence. When $x$ is resolved, the windowed measurement stays rank-1 for every $\delta$.

2-Coherence is lost only when $x$ is also coarse-grained over several lattice points, because then $(-1)^{\sqrt{2}|\beta|x}$ alternates inside the bin and the coherence terms cancel, or when $x$ is ignored (as in parity-only observables).

3-In the continuum approximation, the phase $e^{i\pi\sqrt{2}|\beta|w}$ is treated as a continuously varying function of $\delta$. The integral over the window then averages it out. This produces the decay of $C_\delta^{\rm cont}$ and is an artifact of the approximation.

**Limits.** The argument uses the strong-LO approximation. For finite |β| the Kraus operators for different w at fixed x are only approximately proportional. Sec. 6.3.2 and Fig. 5 quantify this correction for the remote-state purity, using the exact Kraus operators of Eq. (3).

## 4 Numerical Results

### 4.1 Coherence Factor $C_\delta$

Figure 2 compares the coherence factor of the continuum approximation with the discrete strong-LO result. In the continuum approximation, the coherence factor starts at 1 for $\delta = 0$ and decays with oscillations. Its first zero occurs near $\delta \approx 1/(\sqrt{2}|\beta|)$, that is, at $\delta \approx 0.35, 0.14$ and $0.07$ for $|\beta| = 2, 5$ and $10$. Larger $|\beta|$ therefore leads to

faster decay. In contrast, the discrete strong-LO coherence conditioned on $x$ equals 1 for every $\delta$ (Sec. 3.5). The apparent loss of coherence is thus a property of the continuum approximation and not of the windowed measurement itself.

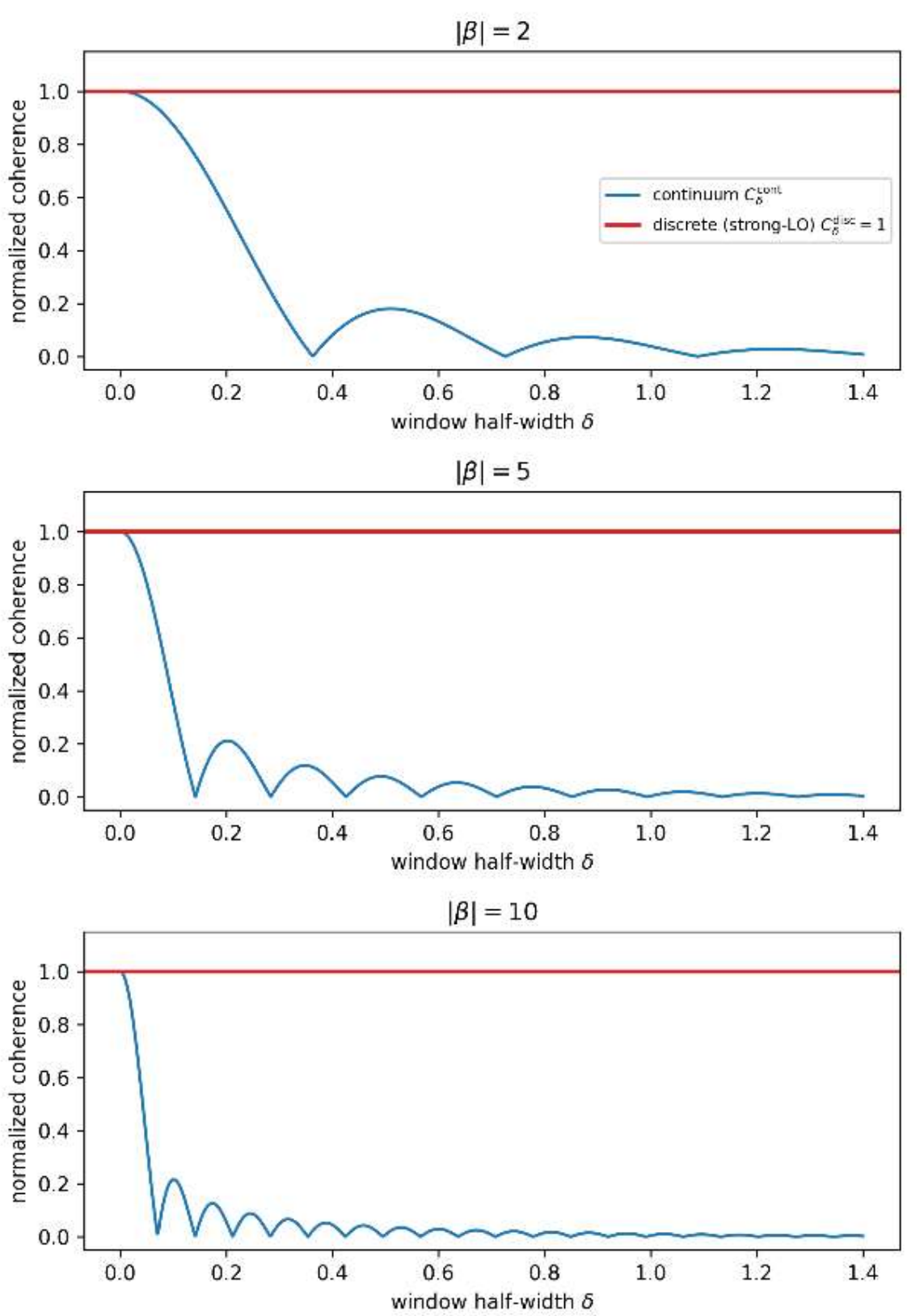

**Fig. 2.** Normalized coherence as a function of the window half-width$\delta$, for $|\beta| = 2,5,10$ (top to bottom). Blue: continuum approximation. Red: discrete strong-LO result, which equals 1 for every $\delta$ when the difference outcome $x$ is resolved (Sec. 3.5).

### 4.2 Parity Signal $D_\delta$

Figure 3 compares the exact parity signal with the continuum prediction. The continuum curve is a smooth oscillation whose first extremum lies near $\delta \approx 1/(2\sqrt{2}|\beta|)$, i.e. $\delta \approx 0.18$, $0.07$ and $0.035$ for $|\beta| = 2$, $5$ and $10$. The exact signal differs from it in three ways.

First, it is piecewise constant. It changes only when an integer value of $n + m$ enters or leaves the window, so the jumps are separated by $1/(\sqrt{2}|\beta|)$ in $\delta$ (for $|\beta| = 5$, at $\delta \approx 0.14,0.28,0.42$).

Second, its sign changes abruptly at these jumps, whereas the continuum curve changes sign smoothly. The signs agree for large |β|, but the exact signal can vanish identically where the continuum value is nonzero; for |β| = 2.3 it is zero for δ below about 0.09.

Third, its magnitude can be much larger than the continuum prediction. For $|\beta| = 2$ and $\delta = 0.05$ the exact signal is $0.39$, whereas the continuum value is $0.11$; for $|\beta| = 2.3$ and $\delta = 0.1$ the exact value is $-0.33$ versus $-0.12$; for $|\beta| = 5$ and $\delta = 0.05$ it is $-0.16$ versus $-0.09$.

For large $|\beta|$ the continuum curve follows the sign changes of the exact signal and can be viewed as a smoothed version of it, while the amplitude remains underestimated.

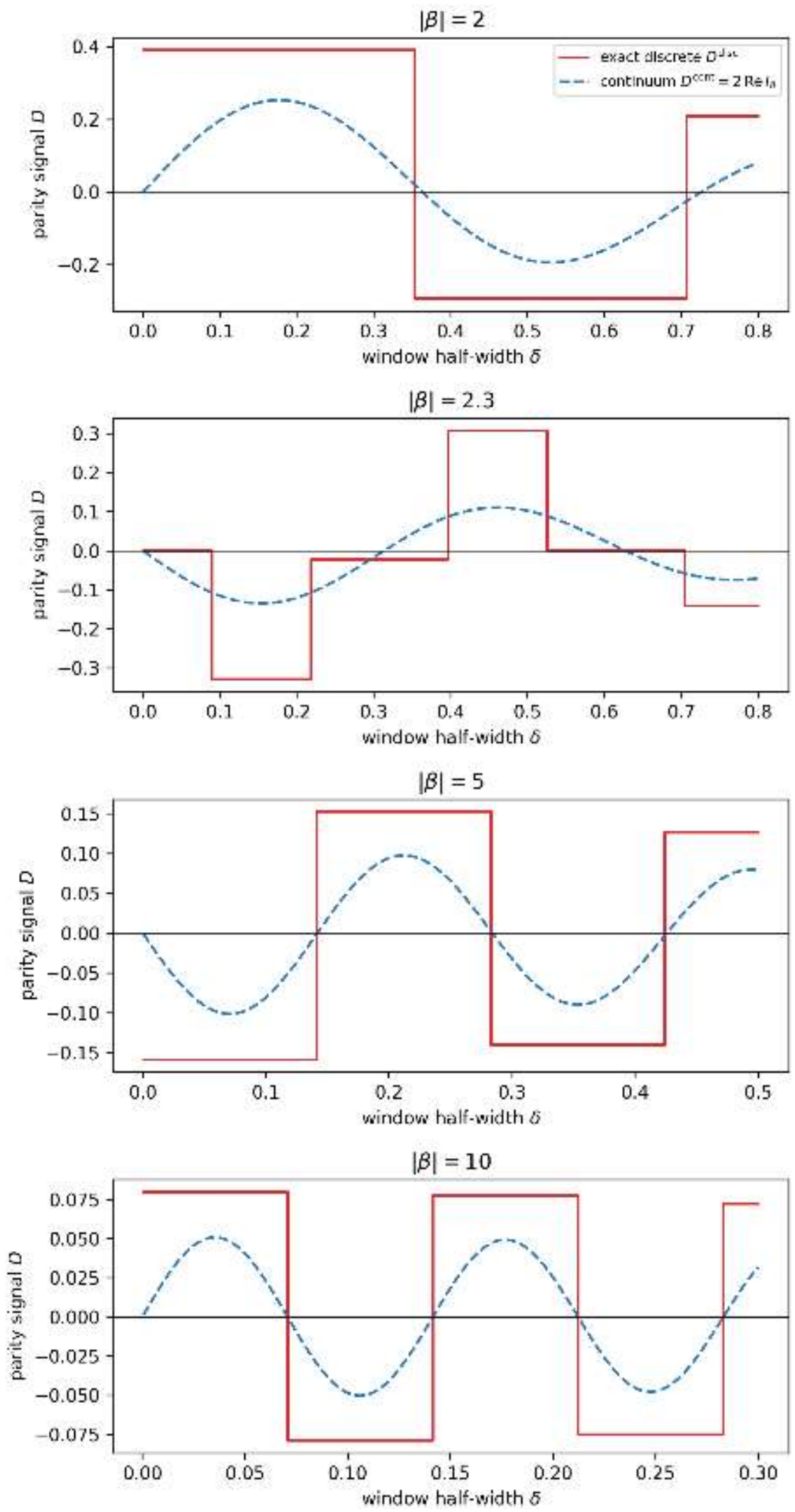

**Fig. 3.** Parity signal $D$ as a function of the window half-width$\delta$, for $|\beta| = 2,2.3,5,10$ (top to bottom). Red: exact discrete signal $D = P(0,+) - P(1,+)$. Blue dashed: continuum approximation $D = 2\,\mathrm{Re}\, I_\delta$. The exact signal is piecewise constant and changes only when an integer value of $n + m$ enters or leaves the window.

## 5 Discrete Parity Structure

### 5.1 Motivation

The analytical results of Sec. 3 are based on the continuum approximation for $w$. In an actual experiment, however, the photon counts $n$ and $m$ are integers, so $w$takes only discrete values. In this section, we show that this discreteness reveals a parity structure that is not captured by the continuum approximation.

### 5.2 Exact Photon-Number Distribution

For vacuum input and a cat-state LO, the exact joint photon-number distribution is [8]

$$P\left(n,m|\beta,\pm,0\right)=\frac{e^{-|\beta|^2}}{n!m!\mathcal{N}_\pm^2(\beta)}\left(\frac{|\beta|^2}{2}\right)^{n+m}\left[(-1)^{m-n}\pm1\right]^2, \tag{23}$$

where $\mathcal{N}_\pm^2(\beta)=2\left(1\pm e^{-2|\beta|^2}\right)$.

The factor $\left[(-1)^{m-n}\pm1\right]^2$ imposes the "parity selection rule": for the even cat, the probability vanishes unless $n+m$ is even; for the odd cat, it vanishes unless $n+m$ is odd.

### 5.3 Windowed success probability

The success probability for the window $[w_\circ-\delta, w_\circ+\delta]$, with $w_\circ=\frac{|\beta|}{\sqrt{2}}$, is obtained by summing the exact distribution over all photon-number pairs whose $w=\frac{n+m}{\sqrt{2}|\beta|}$ lies inside the window:

$$P_{\text{succ}}^{\pm}(\beta,\delta)=\sum_{\substack{n,m\geq0\\ |\beta|^2-\sqrt{2}|\beta|\delta\leq n+m\leq|\beta|^2+\sqrt{2}|\beta|\delta}}P\left(n,m|\beta,\pm,0\right). \tag{24}$$

Defining the integer bounds $s_{\min}=\left[|\beta|^2-\sqrt{2}|\beta|\delta\right]$, and $s_{\max}=\left[|\beta|^2+\sqrt{2}|\beta|\delta\right]$, and using the binomial identity $\sum_{n+m=s}\frac{1}{m!n!}=\frac{2^s}{s!}$, the inner sum can be evaluated explicitly:

$$P(\beta,\delta)=\sum_{s=s_{\min}}^{s_{\max}}\frac{e^{-|\beta|^2}\left(|\beta|^2\right)^s}{\mathcal{N}_\pm^2(\beta)\,s!}\left[(-1)^s\pm1\right]^2. \tag{25}$$

### 5.4 Discrete versus Continuum

Figure 4 compares the exact discrete success probability with two continuum expressions: $G_\delta=\mathrm{erf}(\delta)$, and $G_\delta+\mathrm{Re}\,I_\delta$, which includes the parity term of Eq. (14). For large $|\beta|$ ($|\beta|=5,10$) the steps are small and both continuum curves follow the exact result on average, with $G_\delta+\mathrm{Re}\,I_\delta$ tracking the individual steps more closely.

For small $|\beta|$ the continuum description fails. For $|\beta|=2$, the exact probability is a single plateau of about 0.39 up to $\delta\approx0.71$, followed by a jump to about 0.89; neither continuum curve reproduces this.

For $|\beta|=5$, $|\beta|^2=25$ is odd, so a vacuum input has zero success probability for windows narrower than $\delta\approx0.14$. This parity effect is absent in $G_\delta=\mathrm{erf}(\delta)$ and is only partly captured by $G_\delta+\mathrm{Re}\,I_\delta$.

The steps are separated by $\sqrt{2}/|\beta|$ in $\delta$, since for the even cat only even values of $n+m$ have nonzero probability.

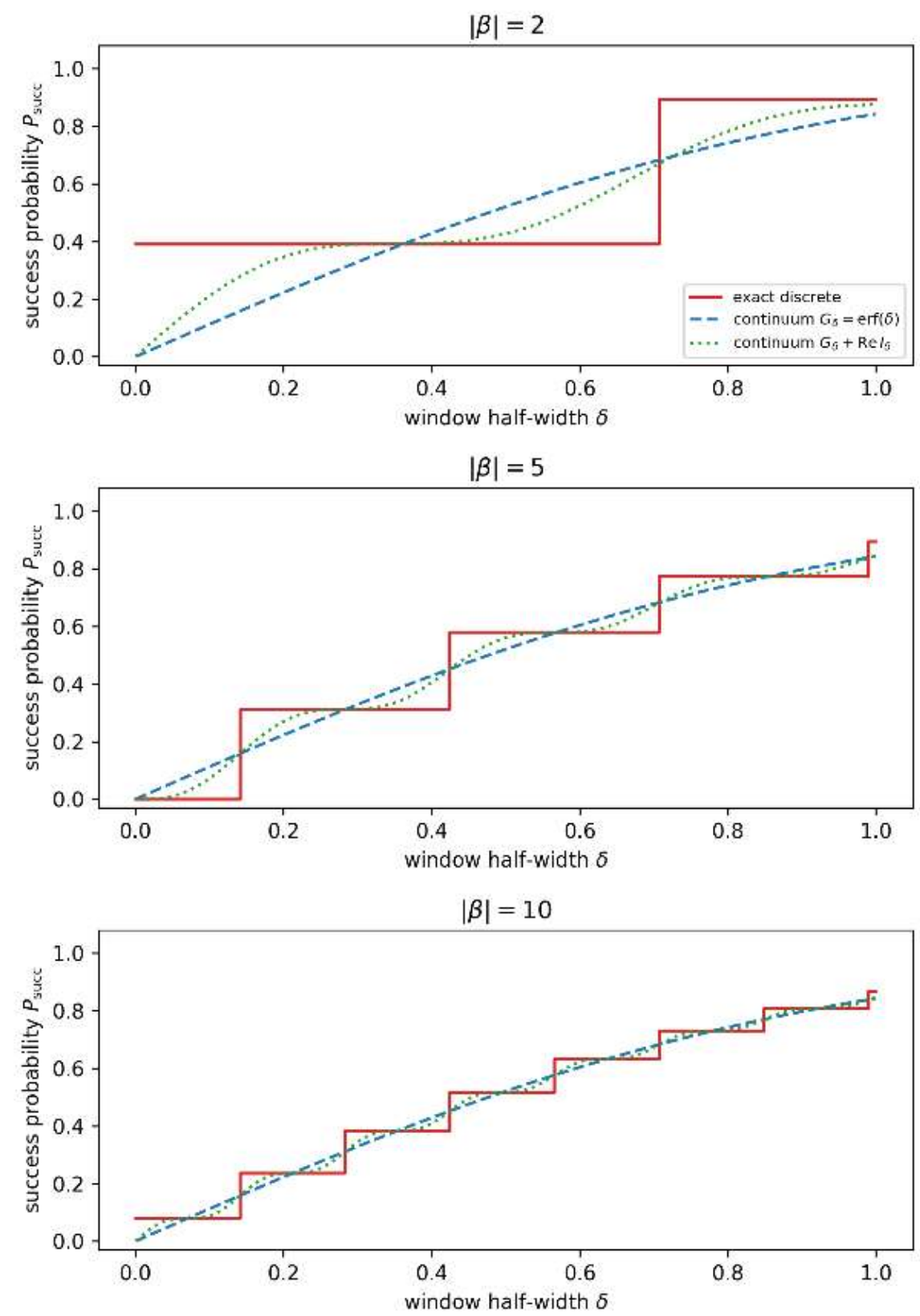

**Fig. 4.** Success probability for a vacuum input and an even-cat local oscillator as a function of the window half-width $\delta$, for $|\beta|=2,5,10$ (top to bottom). Red: exact discrete result. Blue dashed: continuum approximation $G_\delta=\mathrm{erf}(\delta)$. Green dotted: continuum approximation including the parity term, $G_\delta+\mathrm{Re}\,I_\delta$. The exact result is a staircase because only even values of $n+m$ have nonzero probability.

## 6 Discussion

### 6.1 Measurement Engineering with a Windowed Sum Variable

Within the continuum approximation, the window width $\delta$ appears to tune the coherence continuously, from a rank-1 projector (full coherence) to a rank-2 reflection-symmetric POVM (no coherence). On the discrete lattice, however, the coherence conditioned on x is preserved for every window in the strong-LO regime (Sec. 3.5), while the parity-sensitive acceptance is a staircase function of δ and |β|. The window on w can therefore be chosen freely, for example to match the resolution of the detectors, without losing the coherence of the measurement.

The normalized coherence factor $C_\delta^{\text{cont}}$ describes the coherence predicted by the continuum approximation, whereas the discrete strong-LO coherence conditioned on $x$ equals 1. The parity-discrimination signal $D_\delta$ quantifies the sensitivity of the success probability to the parity of the input state. Its exact value is a staircase function of $\delta$ and $|\beta|$ and can be tuned by choosing the window width and the local-oscillator amplitude (Fig. 3).

## 6.2 Comparison with Previous Work

Combes and Lund treated two limiting cases: full retention of the sum variable $w$, which gives a rank-1 POVM with coherence, and complete integration over $w$ in the continuum approximation, which gives a rank-2 reflection-symmetric POVM [8]. They also remarked that the sum-variable information can be used or ignored to engineer a measurement. Our work studies the intermediate case of a finite window and shows, on the discrete lattice in the strong-LO regime, that coherence conditioned on x is preserved for every window.

Thekkadath et al. showed that one can tune continuously between photon-number and quadrature measurements by varying the amplitude of a coherent-state LO [9]. Our setting differs in two respects: the LO is a cat state of fixed amplitude, and the varied quantity is the width of the window applied to the sum variable rather than the LO amplitude.

Olivares et al. studied the role of the LO intensity in homodyne-like tomography with photon-number-resolving detectors [10]. Our analysis is complementary: it concerns the structure of the measurement operators themselves and the role of the sum variable, rather than the reconstruction of states.

## 6.3 Implications for State Preparation

Windowing of the sum variable has operational consequences for remote state preparation (RSP) with catodyne measurements [8]. We discuss the ideal EPR limit in the strong-LO regime.

### 6.3.1. Protocol with an ideal EPR State

Consider an ideal EPR state shared between two parties,

$$|\mathrm{EPR}\rangle_{ab} = \int dx\,|x\rangle_a \otimes |x\rangle_b\,. \tag{26}$$

A catodyne measurement is performed on mode $a$ with outcome $(n,m)$ and we define. $q = (n-m)/\left(\sqrt{2}|\beta|\right)$ The conditional state of mode $b$ is the Schrödinger-cat state in position

$$|c_q\rangle = \frac{1}{\sqrt{2}}\left(|q\rangle + \sigma|-q\rangle\right),\ \sigma = \pm(-1)^{\sqrt{2}|\beta|q} \tag{27}$$

where $\pm$ is the parity of the cat LO. Because $n+m$ and $n-m$ have the same parity, the relative sign $\sigma$ is determined by $q$ alone. It does not depend on the sum variable $w$.

### 6.3.2. Purity of the remote state under windowing

Consider a window $W$ of the sum variable and suppose $q$ is resolved. In the strong-LO limit, all outcomes in the window with the same $q$ give the same relative sign $\sigma$ and the same conditional state up to a global phase (Sec. 3.5). Windowing $w$ then does not mix different states, and the remote state is pure for every window width $\delta$.

More generally, suppose the relative sign is known only through probabilities $p_+$ and $p_-$ for $\sigma = +1$ and $\sigma = -1$. The remote state is then $\rho = p_+|c_+\rangle\langle c_+| + p_-|c_-\rangle\langle c_-|$. Since the two cat states are orthogonal, its linear purity is

$$\mathcal{P} = p_+^2 - p_-^2 = \frac{1+C^2}{2},\ C = |p_+ - p_-|. \tag{28}$$

For a windowed measurement with q resolved, $p_+ = 1$ or $p_- = 1$, so $C = 1$ and $\mathcal{P} = 1$.

Exact calculation at finite $|\beta|$. To go beyond the strong-LO limit, we computed the remote state with the exact Kraus operators of Eq. (3). We used a two-mode squeezed vacuum $\propto \sum_k \lambda^k |k\rangle|k\rangle$ with $\lambda = \tanh r$, which becomes the ideal EPR state for $\lambda \to 1$. Here $\lambda = 0.882$ for 12 dB and $\lambda = 0.6$ for 6 dB of squeezing. For an outcome $(n, m)$ the unnormalized remote state is $\sum_k \lambda^k \langle 0|M_{n,m}|k\rangle\,|k\rangle$. When x is resolved and only w is windowed, the remote state is the mixture of the states with the same $n-m$ and with $n+m$ in the window, weighted by their probabilities. We evaluate its purity $\mathcal{P} = \sum_{ij} p_i p_j \left|\langle\psi_i|\psi_j\rangle\right|^2$.

Figure 5 (top) shows the result for 12 dB of squeezing. With x resolved, the purity decreases slowly with $\delta$ and approaches 1 as $|\beta|$ grows. For $\delta = 1$ we find $\mathcal{P} = 0.964,\ 0.970,\ 0.986,\ 0.993$ for $|\beta| = 3,5,8,12$. For 6 dB of squeezing, $\mathcal{P} \geq 0.98$ for all $|\beta| \geq 2$ up to $\delta = 1$. The window therefore acts as a weak control of purity at finite $|\beta|$, and not as a switch between a pure and a mixed state. When x is not recorded (Fig. 5, bottom), the purity drops to about 0.15 to 0.55 for $|\beta| = 2$ to 5, because states with different q and parity are mixed.

For the ideal EPR state ($\lambda \to 1$) the purity does not approach 1 as $|\beta|$ grows at fixed $\delta$ (for $\lambda = 0.99$ we find $\mathcal{P} \approx 0.97$ and 0.90 at $\delta = 0.5$ and 1 for $|\beta| = 12$). This is expected, since the ideal EPR state has unbounded photon number, for which the strong-LO approximation is not uniformly valid [8]. Finite squeezing is therefore essential for the conclusions above.

Limitations: we assumed ideal photon-number-resolving detectors, a single mode, and a two-mode squeezed vacuum as the resource. The practical consequence is positive: the window on w can be chosen freely, for instance to match a finite detector resolution in the sum variable, with only a small loss of purity of the prepared cat state, provided x is resolved. This is relevant for error-correcting codes based on cat states [13, 14].

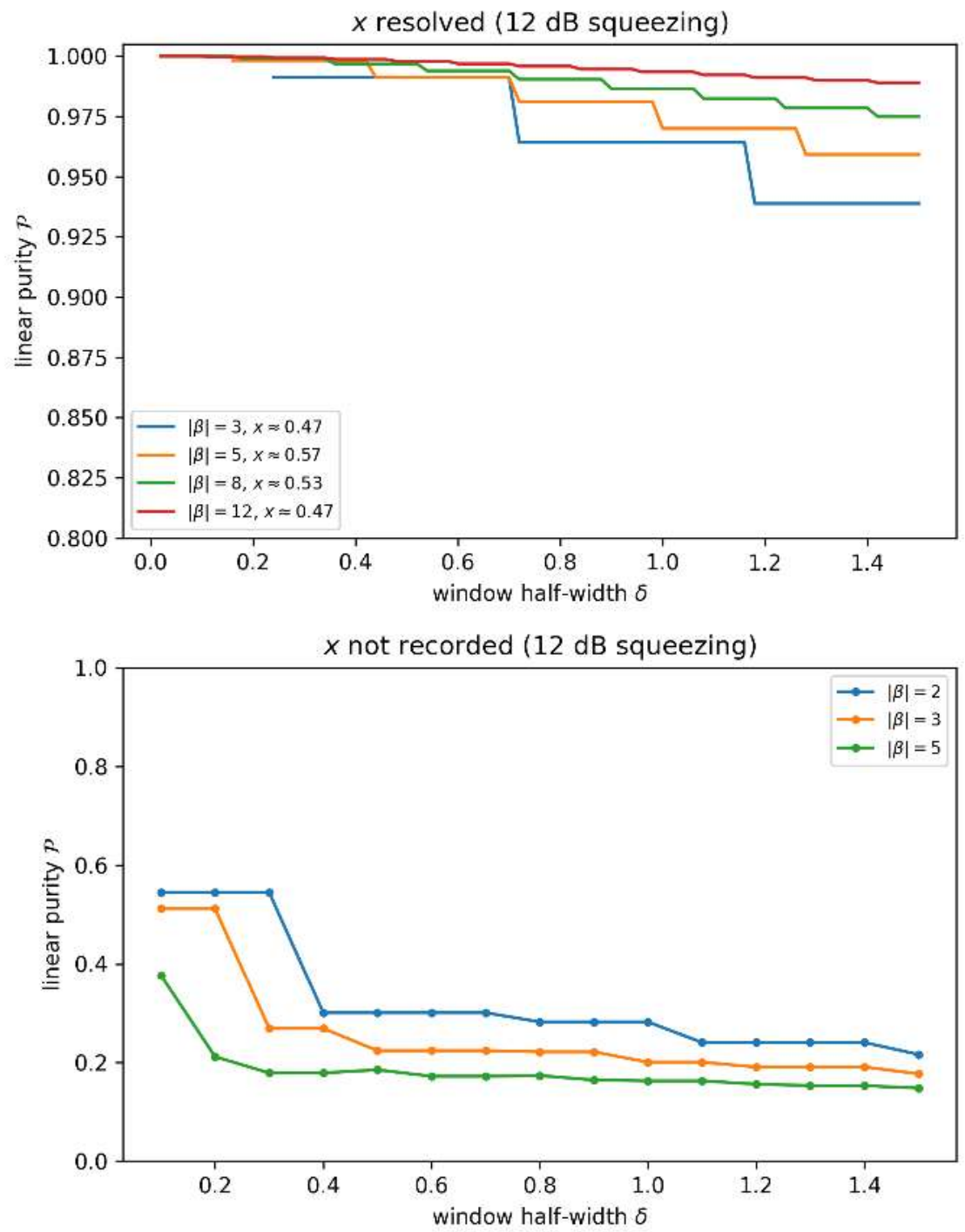

**Fig. 5.** Linear purity of the remote state as a function of the window half-width $\delta$, for 12 dB of two-mode squeezing, computed with the exact Kraus operators. Top: $x$ resolved, for $|\beta| = 3,5,8,12$ (with $x \approx 0.5$). Bottom: $x$ not recorded, for $|\beta| = 2,3,5$.

### 6.4 Experimental Feasibility

Realizing the windowed measurement requires photon-number-resolving detectors that resolve single photons in both outputs, because the exact windowed quantities depend on the integer value of n + m (Secs. 3.5 and 5). The window width in photon-number units is $2\sqrt{2}|\beta|\delta$. For $|\beta| = 5$ and $\delta = 0.1$ it is about 1.4, so the window contains only one or two integer values of n + m. In this regime the continuum formulas are not valid and the exact discrete expressions must be used.

Current detectors can resolve more than 15 photons with reasonable efficiency [11]. This is at the edge of what is needed for $|\beta| = 5$, where each detector registers on average 12.5 photons. A cat state with $|\beta| = 5$ (mean photon number 25) is also well beyond current optical experiments. A smaller amplitude such as $|\beta| = 2$ (mean photon number 4) is more realistic, and it is also the regime where the discrete and continuum descriptions differ most. In the microwave domain, photon-number resolution inside a cavity is well established [12], and Combes and Lund consider a realization there plausible [8].

Limitations. Our analysis assumes ideal detectors. The parity structure is encoded in outcomes that differ by a single photon, so photon loss and detector inefficiency, which shift n or m by one, reduce the visibility of the parity signal and flip the relative sign $\sigma$ of the remote cat state [8]. Their effect should be quantified in future work. Furthermore, our analysis is restricted to single-mode fields; an extension to multimode fields is needed for practical applications.

## 7 Conclusions

We analyzed catodyne measurements in which only a finite window of the photon-sum variable is retained. Our main results are:

**1-Coherence conditioned on x is preserved in the strong-LO regime.** On the discrete photon-number lattice, windowing the sum variable does not reduce the coherence conditioned on the difference outcome. The apparent decay of coherence in the continuum approximation is an artifact of treating the sum variable as continuous.

**2-Parity-sensitive acceptance.** The exact windowed success probability depends on the parity of the input state. The parity signal is a staircase function of the window width, and its magnitude and plateau structure differ substantially from the continuum prediction.

**3-Limits of the continuum approximation.** For the success probability, the continuum formulas reproduce the average behavior only for large $|\beta|$; for small $|\beta|$ and narrow windows, which are the relevant regime for photon-number-resolving detectors, the discrete structure must be used. The amplitude of the parity signal is underestimated by the continuum formula even for $|\beta| = 10$.

For remote state preparation, the conditional cat state keeps a purity close to 1 when $x$ is resolved (computed exactly for finite squeezing), and loses purity mainly when $x$ is not recorded. Our analysis assumes ideal detectors and a single mode; detector inefficiency and multimode effects remain to be studied.

## References

[1] T. Tyc and B. C. Sanders, Operational formulation of homodyne detection, J. Phys. A: Math. Gen. **37**, 7341 (2004).

[2] R. Kumar, E. Barrios, A. MacRae, E. Cairns, E. H. Huntington, and A. I. Lvovsky, Versatile wideband balanced detector for quantum optical homodyne tomography, Opt. Commun. **285**, 5259 (2012).

[3] N. Walker, Quantum theory of multiport optical homodyning, J. Mod. Opt. **34**, 15 (1987).

[4] S. L. Braunstein and P. van Loock, Quantum information with continuous variables, Rev. Mod. Phys. **77**, 513 (2005).

[5] B. C. Sanders, K. S. Lee, and M. S. Kim, Optical homodyne measurements and entangled coherent states, Phys. Rev. A. **52**, 735 (1995).

[6] G. S. Thekkadath, B. A. Bell, I. A. Walmsley, and A. I. Lvovsky, Engineering Schrödinger cat states with

a photonic even-parity detector, Quantum. **4**, 239 (2020).

[7] M. S. Kim and B. C. Sanders, Squeezing and antisqueezing in homodyne measurements, Phys. Rev. A. **53**, 3694 (1996).

[8] J. Combes and A. P. Lund, Homodyne measurement with a Schrödinger cat state as a local oscillator, Phys. Rev. A. **106**, 063706 (2022).

[9] G. S. Thekkadath, D. S. Phillips, J. F. F. Bulmer, W. R. Clements, A. Eckstein, B. A. Bell, J. Lugani, T. A. W. Wolterink, A. Lita, S. W. Nam, T. Gerrits, C. G. Wade, and I. A. Walmsley, Tuning between photon-number and quadrature measurements with weak-field homodyne detection, Phys. Rev. A.**101**, 031801(R) (2020).

[10] S. Olivares, A. Allevi, and M. Bondani, On the role of the local oscillator intensity in optical homodyne-like tomography, Phys. Lett. A. **384**, 126354 (2020).

[11] L. A. Morais, T. Weinhold, M. P. de Almeida, J. Combes, A. Lita, T. Gerrits, S. W. Nam, A. G. White, and G. Gillett, Precisely determining photon-number in real-time, arXiv: 2012.10158.

[12] D. I. Schuster, A. A. Houck, J. A. Schreier, A. Wallraff, J. M. Gambetta, A. Blais, L. Frunzio, J. Majer, B. Johnson, M. H. Devoret, S. M. Girvin, and R. J. Schoelkopf, Resolving photon number states in a superconducting circuit, Nature. **445**, 515 (2007).

[13] D. S. Schlegel, F. Minganti, and V. Savona, Quantum error correction using squeezed Schrödinger cat states, Phys. Rev. A. **106**, 022431 (2022).

[14] Q. Xu, G. Zheng, Y.-X. Wang, P. Zoller, A. A. Clerk, and L. Jiang, Autonomous quantum error correction and fault-tolerant quantum computation with squeezed cat qubits, arXiv: 2210.13406.